# STATUS AND CONTROLS REQUIREMENTS OF THE PLANNED HEAVY ION TUMOR THERAPY ACCELERATOR FACILITY HICAT

R. Bär, H. Eickhoff, T. Haberer; GSI Darmstadt, Germany

## Abstract

The HICAT project is a Heavy Ion accelerator for light ion Cancer Treatment to be built for the clinics in Heidelberg, Germany. It consists of a 7 MeV/u linac, a compact synchrotron and three treatment places, one of them located after a 360 degree gantry beam-line. The facility will implement the intensity controlled raster-scanning technique that was developed and successfully demonstrated at GSI. In order to produce the beams with the characteristics requested by the treatment sequencer, the accelerator must operate on a pulse-to-pulse basis with different settings. This concept imposes strict and challenging demands on the operation of the accelerators and hence the control system of the facility. The control system should be developed, installed and maintained by and under the complete responsibility of an industrial system provider, using a state-of-the-art system and wide-spread industrial components wherever possible. This presentation covers the status of the project and the requirements on the control system.

## 1 HICAT MACHINE LAYOUT

The HICAT accelerator facility was designed to meet the medical requirements used for the successful GSI cancer treatment program. HICAT consists of two ECR ion sources with independent spectrometer lines to produce both low LET ions (p, He) and high LET ions (C, O) to cover the medical requirements and allow for fast switching. The beams will be accelerated by a compact-designed 7 MeV/u linac (RFQ and IH-DTL) and a 6.5 Tm synchrotron for beam energies of up to 430 MeV/u. In order to meet the intensity requirements, multi-turn injection will be used for beam accumulation. For extraction, the rf knock-out method in combination with a variable extraction time will be implemented. The beams can be delivered to three medical treatment areas, located after two fixed horizontal beam lines and one 360° rotating beam transport system (isocentric gantry), as well as one quality assurance/experimental place.

No passive elements are foreseen to match the dose distribution to the individual tumor geometry. Instead, all beam lines will be equipped with horizontal and vertical scanning magnets and beam diagnostic devices for the intensity controlled-raster scanning technique that was developed and successfully demonstrated within the GSI experimental cancer treatment program with over 100 patients to date.

Fig. 1 shows the cross section of the first underground floor of the building which houses the accelerator sections, the patient treatment areas, local control rooms and various laboratories and offices. For a general description of the accelerator complex, see [1].

## 2 CONTROL SYSTEM

For the realization of the HICAT project, a close cooperation with industry is necessary. GSI has done the machine layout and worked out crucial technical specifications. The development and implementation of the control system will be performed by a commercial firm.

Therefore, the HICAT control system (cs) should be delivered under a supply contract established with an industrial system provider. The full development and configuration of the system will be the responsibility of the contractor, who will also perform commissioning and all tests on the site. GSI will advise and supervise the commercial contractor during the project.

The cs should be based on an industrial control system (e.g. commercially available SCADA system) that fulfills the specific requirements for controlling the facility. At a life cycle of 20 or more years, quite reasonable for such a system, the cs must be laid out for long service. Consequently, it is proposed to use wide-spread industrial standards and commercial solutions wherever possible.

The cs must meet the following general requirements:

- high reliability
- easy and efficient operation by a small crew
- development and maintenance by industry

Since a reliable control system is an indispensable element to supply beams for medical practice the system has to respect stringent requirements in terms of reliability, safety, maintainability and legal aspects such as the ones required by regulatory authorities. It must meet the following general requirements:

The HICAT cs runs the accelerator in a hospital environment. It must be easy to operate because daily operation of the accelerator is carried out by operators who are not necessarily accelerator specialists. The control system's functionality is determined by these facts, together with the controls demands of the accelerator facility in itself.

In particular, the following major tasks have to be covered by the accelerator cs:

- full control of the accelerator systems to deliver and assure a beam with characteristics requested by the patients' treatment plan on a pulse to pulse basis
- full control of the patient irradiation treatment with prescribed treatment plan and measure compliance of treatment
- patients' protection against wrong treatment
- control of all secondary and infrastructure systems
- minimal personal employment for controlling

The HICAT cs should be completely integrated, i.e. a single set of standards will be applied to all aspects of the facility. Although it is not the usual practice to include conventional facility controls in the accelerator cs, experience shows that signals from these systems are needed in the control room.

All systems – accelerators and conventional facilities – will be operated and monitored from a single control room. In addition there will be treatment control rooms for each treatment place from where the actual treatment is controlled and monitored.

## 3 TECHNICAL CONTROL REQUIREMENTS

Major aspects of the control system design are influenced by the experiences of the GSI cancer treatment program [2]; the requirements of this facility, however, exceeds those of the pilot project.

Logically, the HICAT cs will consist of two major integrated parts:

- the accelerator control system – controls the accelerator and all auxiliary systems
- control system for the medical treatment (one independent system for each treatment place) – controls the treatment sequencer and all treatment instrumentation

### 3.1 Process Data

The main characteristics of the HICAT facility are the application of the raster-scan method with active intensity-, energy- and beam size-variation in combination with the usage of an isocentric gantry to match the dose distribution to the individual tumor geometry. During the patient irradiation process (treatment mode) the treatment sequencer requests beams with certain characteristics on a pulse-to-pulse basis. The beam characteristics are fixed and pre-defined in a library of settings, e.g. 255 possible beam energy steps (Table 1). The task of the accelerator control system is to deliver these beams as requested.

Table 1: Process parameters

| Steps | Parameter |
|---|---|
| 4 | ion species (p, He, C, O) |
| 255 | beam energy (50–430 MeV/u) |
| 4 | focus diameter (4–10 mm) |
| 15 | intensity level (variation 1 to $10^{-3}$) |
| ~36 | gantry angle (angle interpolation) |
| 4 | treatment area (beam lines) |

In order to achieve high operational reliability the cs must meet the following requirements:

- use only proven and validated settings,
- protect settings from loss and accidental modification,
- inhibit operator's access/interference during irradiation,
- avoid networking for settings, provide all beams in stand-by on a pulse-to-pulse basis,

These requirements lead to a high number of settings (up to 100,000 per device) to hold at the device controller. Fortunately, the settings of the devices do not depend on all parameters in Table 1 at the same time. Because of the high number of data sets needed, manual tuning of the machine is prohibitive. To derive data automatically, a "theoretical model" of the accelerator will be used, allowing calculation of all device data from high-level machine parameters and applying once measured corrections.

### 3.2 Timing system

For the entire accelerator chain, including linac rf-structures, choppers, bunchers, and all synchrotron devices, a strict synchronization is required. Some time-critical functions (i.e. triggers, acceleration ramps) must by synchronized within $\Delta t < 1$–5 µs at the worst. The cs must implement a timing controller that provides a system-wide synchronization for distributed devices with real-time performance. The system must be flexible enough to react on requests for beam extraction pauses and aborts.

## 3.3 Patient safety

Safety is a crucial issue for the cs, considering the special task of the facility. Due to the high number of devices acting on the beam, the accelerator control system cannot guarantee safe operation. Instead, it will implement a fast and fail-safe beam-abort system that reliably stops the beam within 200 µs. This is achieved by simultaneously triggering a chain of different actions using different devices which are all, by themselves, capable to prevent beam delivery to the treatment place.

The sophisticated on-line beam diagnostics of the treatment sequencer verifies the correctness of the beam parameters (position, focus, intensity) with respect to the treatment plan. In case of intolerable deviations to the treatment plan or failure of critical systems, the treatment control system triggers beam-abort channels to stop the irradiation treatment.

## 3.4 Scanner system

For the real-time treatment sequencer it is foreseen to use the existing state-of-the-art GSI-system [3] and adapt it to the new environment. The functions will be extended to feature gated beam extraction and end of extraction ahead of schedule.

## 4 STATUS AND OUTLOOK

Currently all essential technical specifications and requirements for the entire facility including the control system are being worked out.

Following the final approval by the board of the clinics by the end of 2001 an ambitious time schedule will be established. For the first quarter of 2002 a call for tender is scheduled. Contracts with industries are foreseen around middle of 2002.

According to the current project plan the first beam is scheduled in 2005 and the first patient treatments should take place in 2006 after an extensive commissioning phase.

## REFERENCES

[1] H. Eickhoff *et. al.*, "The Proposed Dedicated Ion Beam Facility for Cancer Therapy at the Clinic in Heidelberg", EPAC2000, Vienna, June 2000.
[2] U. Krause, R. Steiner, "Adaption of a Synchrotron Control System for Heavy Ion Tumor Therapy", ICALEPCS '95, Chicago, USA, 1995.
[3] E. Badura *et. al.*, "Safety and Control System for the GSI Theray Project", ICALEPCS '97, Beijing, China, November 1997.

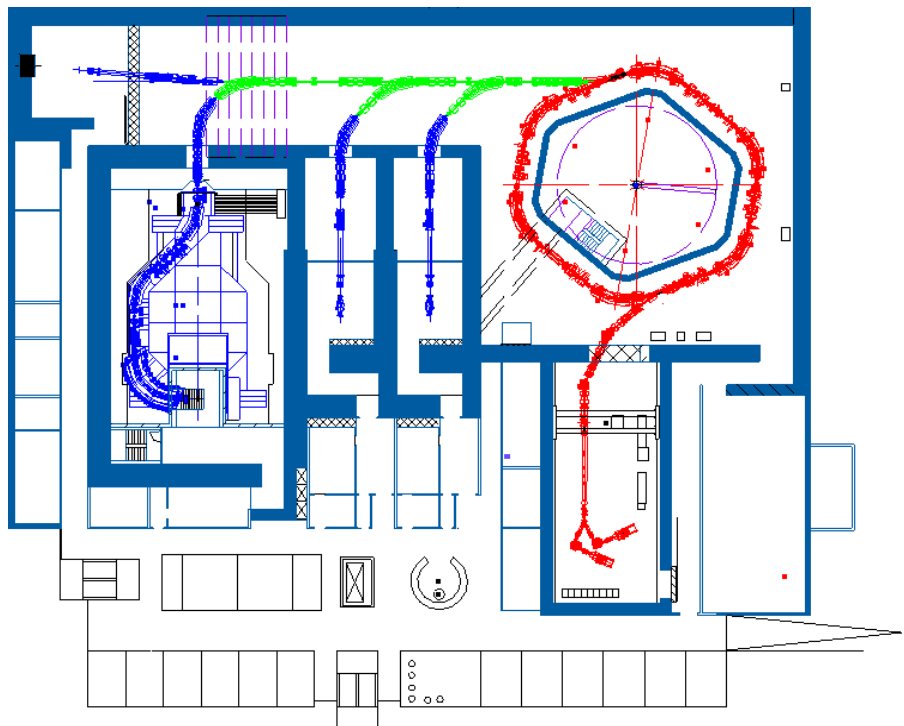

Figure 1: First underground level of the HICAT facility.